\documentclass[useAMS,usenatbib]{mn2e}
\usepackage{graphicx}

\title[High-speed photometry of faint CVs]{High-speed photometry of faint cataclysmic variables - VIII. Targets from the Catalina Real-time Transient Survey}
\author[Deanne L. Coppejans et al.]{Deanne L. Coppejans$^{1,2}$\thanks{Email: d.debude@astro.ru.nl}, Patrick A. Woudt$^{2}$, Brian Warner$^{2,3}$, Elmar K\"{o}rding$^1$,\newauthor Sally A. Macfarlane$^{1,2}$, Matthew P.E. Schurch$^{2}$, Marissa M. Kotze$^{4,2}$,
\newauthor Hannes B. Breytenbach$^{2}$, Amanda A. S. Gulbis$^{4,5,6}$ and Rocco Coppejans$^{4,1,2}$\\
$^{1}$Department of Astrophysics/IMAPP, Radboud University Nijmegen, P.O. Box 9010, 6500 GL Nijmegen, The Netherlands\\
$^{2}$Astrophysics, Cosmology and Gravity Centre, Department of Astronomy, University of Cape Town, Private Bag X3,\\Rondebosch 7701, South Africa\\
$^{3}$School of Physics and Astronomy, Southampton University, Highfield, Southampton SO17 1BJ, UK\\
$^{4}$South African Astronomical Observatory, P.O. Box 9, Observatory, 7935, Cape Town, South Africa\\
$^{5}$Southern African Large Telescope, P.O. Box 9, Observatory, 7935, Cape Town, South Africa\\
$^{6}$Department of Earth, Atmospheric, and Planetary Sciences, Massachusetts Institute of Technology, 77 Massachusetts Avenue,\\Cambridge, MA 02139-4307, USA}
\begin{document}

\date{}

\pagerange{\pageref{firstpage}--\pageref{lastpage}} \pubyear{2013}

\maketitle

\label{firstpage}

\begin{abstract}

Time series photometry of 20 Cataclysmic Variables detected by the Catalina Real-time Transient Survey is presented. 14 of these systems have not been observed previously and only two have been examined in-depth. From the observations we determined 12 new orbital periods and independently found a further two. Eight of the CVs are eclipsing systems, five of which have eclipse depths of more than 0.9 mag. Included in the sample are six SU UMa systems (three of which show superhumps in our photometry), a polar (SSS1944-42) and one system (CSS1417-18) that displays an abnormally fast decline from outburst.
 
\end{abstract}


\section{Introduction}

We present the latest results of a photometric follow-up study of faint Cataclysmic Variable stars (CVs; see \citealt{Warner1995}) mostly in the southern hemisphere. Specifically, in this paper we discuss the observations of 20 CVs identified by the Catalina Real-time Transient Survey (CRTS; see \citealt{Drake2009}). 

This work forms part of a survey with the aim of characterising newly identified CVs, determining their orbital periods, searching for sub-orbital periodicities and selecting targets for in-depth studies on large telescopes and multi-wavelength campaigns. In the previous papers in this series (see \citealt{Woudt2012} and references therein), the focus was on faint nova remnants and CVs identified by the Sloan Digital Sky Survey (SDSS, e.g. \citealt{Aihara2011}). Recently, it has shifted to CVs discovered by the CRTS \citep{Woudt2012}.

The CRTS is a large scale transient survey that observes 30$\,$000 deg$^2$ of the sky in search of transients (see \citealt{Djorgovski2011}). They make use of data from the Catalina Sky Survey \citep{Christensen2012}, which searches for Near-Earth Objects. The observing strategy is to observe a field four times at 10-minute intervals, then return to the field up to four times per lunation \citep{Djorgovski2011}. Transients are detected by looking for variations of more than 2 mag in the $V$ filter. As most of the fields have 7- to 8-year baselines and the CRTS can reach a depth of $V\sim$23 by co-adding images (although individual pointings reach $V\sim$19 to 21), a variety of transients are discovered. Amongst these are supernovae, blazars, flare stars and CVs.

As evidenced by the number of CVs the CRTS has detected - more than 1000 to date (September 2013)\footnote{See http://nesssi.cacr.caltech.edu/catalina/Stats.html} - the CRTS is particularly efficient at finding them. This is for a number of reasons. For example, the magnitude variation cut-off limit of 2 mag, which was designed to restrict the number of artifacts and pulsating variables, is approximately the minimum amplitude of a dwarf nova outburst, so large numbers of Dwarf Novae (DNe) are detected. This range also ensures that the high inclination systems with deep eclipses are detected, as well as a number of magnetic systems that show high and low states. As the CRTS commonly reaches down to $V\sim$21, or fainter if the object has a high state brighter than this limit, it will find faint CVs such as the double degenerate AM CVn (AM Canum Venaticorum) systems. Also advantageous to our photometric follow-up study, is the excellent coverage in the southern hemisphere.

Follow-up observations on the objects identified by large scale surveys, such as the CRTS, are becoming increasingly necessary as existing projects continue to detect new transients and surveys such as the Large Synoptic Survey Telescope (LSST; \citealt{Sweeney2009}) are planned. On average, from 2008 to 2012 the CRTS detected 45 new CVs per quarter. For CVs, finding the orbital periods for a greater sample of systems is necessary for evolutionary studies (e.g. \citealt{Gaensicke2009}). Additionally, amongst the newly discovered CVs will be systems that need to be observed with larger telescopes (e.g., SALT: Woudt et al. 2010) and multi-wavelength campaigns (e.g. CC Sculptoris: Woudt et al. 2012b). Particularly important are the eclipsing systems, from which we can obtain system characteristics such as the mass ratio through eclipse deconvolution (Littlefair et al. 2008; Savoury et al. 2011). Photometric surveys, such as the kind presented in this paper, help to identify these targets.
 
In this paper, in section \ref{sec:observations} we describe the observing procedure, the instrumentation and the data reduction. The results of each of the individual CVs follow. Section \ref{sec:conclusions} contains a summary and discussion of the results.

\section[]{Observations}\label{sec:observations}

Differential photometry was performed using the University of Cape Town (UCT) CCD \citep{ODonoghue1995} and the new Sutherland High-speed Optical Cameras (SHOC; \citealt{Gulbis2011meeting}, \citealt{Coppejans2013}) mounted on the 74-in and 40-in reflector telescopes of the South African Astronomical Observatory (SAAO).

For the first time in this series of papers, we have used the SHOC systems. Two instruments utilizing Andor iXon 888 cameras are available for use at the SAAO. The field of view is 2.85x2.85 arcmin$^2$ on the 40-in telescope and 1.29x1.29 arcmin$^2$ on the 74-in\footnote{See http://shoc.saao.ac.za/}, as compared to 1.23x1.82 arcmin$^2$ and 0.57x0.83 arcmin$^2$ on the UCT CCD.

For these observations both the UCT CCD and SHOC were used in frame-transfer mode and SHOC was operated in 1MHz conventional mode with 2.4 preamplifier gain. No filters were used, but the data were calibrated to the Sloan $r$ photometric system. Papers I to VI in this series used hot white dwarfs as calibration standards \citep{Landolt1992}, which gave $V$ accurate to $\sim$0.1 mag \citep{Woudt2012}. Using the survey archival data we compared the calibration offset using hot white dwarfs and those using SDSS photometry of comparison stars on the target field on nights when both were available. Over the range $g-r$=0.2 to 1.0, it was found that there was a stable zero-point offset of 0.12$\pm$0.05 mag between $V$ and SDSS $r$ - which is consistent with the photometric transformation in \citet{Jester2005}. Subsequently, we calibrated our white light photometry to the $r$ photometric system using SDSS stars on the target field that have $g-r$ colours in the range 0.2-1.0. The calibration is accurate to $\sim$0.1 mag. In the cases where a suitable calibration star was not observed over the course of the evening, we applied the calibration determined for another night during the observing run. The runs calibrated in this way are labelled with a colon in the observing log (Table \ref{tbl:log}).

The observing procedure varies according to the behaviour of a given CV. A target is initially observed for a few hours ($\sim$4 h). If the light curve shows modulation, then it is observed further over consecutive nights. If there are no modulations present or the CV appears to have a long orbital period (greater than 5 hours), then no further observations are taken as the telescope time can be spent more profitably. Longer-term monitoring (weeks to years) is carried out in order to reduce the error margin on orbital periods or to observe CVs in an alternate outburst state in order to obtain a superhump period, an orbital period or to examine Quasi-Periodic Oscillations (QPOs) and Dwarf Nova Oscillations (DNOs).  

Data reductions were performed with the program Duphot for the UCT CCD \citep{ODonoghue1995} and in IRAF \citep{Tody1986} using standard reduction routines for the SHOC observations.

Two different techniques were used to find periodicities (such as the orbital period) in the photometry. The CVs which had sinusoidal light curves were analyzed by Fourier Analysis using the Starlink Period package\footnote{http://www.starlink.rl.ac.uk/docs/sun167.htx/sun167.html}. Those with non-sinusoidal light curves (e.g. the eclipsing systems) were analyzed by Phase Dispersion Minimization (PDM, \citealt{Stellingwerf1978}). The former technique has been explained in detail in previous papers in this series. Briefly, PDM determines the true period by folding the light curve on a series of test periods and determining which produces the least scatter. In order to quantify the scatter, each folded light curve is binned and the overall variance of all the bins is divided by the variance of the unbinned data. The result - the PDM statistic $\Theta$ - will be close to 1 for false periods and be small for true periods. A PDM periodogram will thus have local minima at the true periods (see for example Figure \ref{fig:CSS0116_pdm}). Full details are given in \citet{Stellingwerf1978}. 

The distinction in treatment between the non-sinusoidal and sinusoidal systems is due to the fact that in attempting to fit sharp features by a series of sinusoids, Fourier Analysis spreads the power to harmonics of the period, thereby reducing the power of the fundamental. PDM does not suffer from this problem.

Unless otherwise stated, before the PDM periodogram or Fourier Transform (FT) is calculated, the individual runs are linearly detrended and mean-subtracted. This reduces the spurious noise at low frequencies introduced by observing the CV over the same airmass range over consecutive nights.

In order to determine an uncertainty on the periods obtained from the FT, a sine-curve is fitted to the photometry using the FT period as an initial guess. By varying the period, a Markov Chain Monte Carlo Method (MCMC) - specifically the Metropolis Hastings algorithm (\citealt{Metropolis1953} and \citealt{Hastings1970}) - then samples the probability distribution of solutions. The median value and standard deviation of this distribution are then quoted as the period and uncertainty respectively. In cases where there are additional strong peaks in the FT, more than one sine-wave is fitted simultaneously.

For those systems which have highly non-sinusoidal light curves, the uncertainty cannot be determined in this way. The uncertainty derived from PDM for these CVs is obtained by bootstrapping using the Monte Carlo Algorithm for case resampling \citep{Efron1979}. The sampling distribution for the period is estimated by creating new light curves by selecting N of the N data points (repetition is allowed) and performing PDM on each new sample. As is the case for the FT, the period and uncertainty given in this paper are the median and standard deviation of the bootstrap distribution.

A variety of other techniques are used to select the correct period from the FT/PDM periodogram and determine the reality of a peak. These include phase-folding, using the window function - which shows the aliasing structure introduced by sampling - and the Fisher randomisation test \citep{Nemec1985}. In the latter, the amplitude values on the light curves are shuffled to create a large number of new light curves and FTs are calculated for each. The probability of a period being real and equal to the quoted value on the original FT is then determined by looking at the proportion of shuffled light curves that gave a higher peak at that frequency.

The observing log is given in Tab. 1. Each CRTS transient has an ID of the form CSS yymmdd:hhmmss$\pm$ddmmss. The first three letters indicate in which of the three component surveys it was discovered (CSS: Catalina Sky Survey, MLS: Mount Lemmon Survey and SSS: Siding
Springs Survey, see \citealt{Djorgovski2011}). The following six digits indicate the date on which it was classified. The Right Ascension and Declination follow the colon. In the observing log and the discussion the ID is abbreviated to the form CSS hhmm$\pm$dd.

\begin{table*}\label{tbl:log}
 \centering
 \begin{minipage}{140mm}
  \caption{Observing log}
  \begin{tabular}{lllllllll}
  \hline
  Object & Type & Run No. & Date of obs. & HJD of first obs. & length & t$_{in}$ & Tel. & $r$\\
  & & & (start of night) & (+2450000.0) & (h) & (s) & & (mag)\\
  \hline
  \textbf{CSS0116+09} & DN & S8018$^o$ & 27 November 2010 & 5528.2811 & 2.3485 & 30 & 40-in & 16.8$^m$\\
  && S8021$^o$ & 28 November 2010 & 5529.2925 & 2.3928 & 30 & 40-in & 17.1$^m$\\
  && S8024$^o$ & 29 November 2010 & 5530.2755 & 3.5503 & 15 & 40-in & 17.3$^m$\\
  && S8033 & 5 December 2010 & 5536.2887 & 1.1579 & 20,25 & 74-in & 18.9$^m$\\
  && S8040 & 8 December 2010 & 5539.3026 & 2.0642 & 25 & 74-in & 19.1$^m$\\
  && S8098$^s$ & 5 October 2011 & 5840.4774 & 2.8041 & 20 & 74-in & 18.8$^m$\\
  && S8100$^s$ & 6 October 2011 & 5841.4096 & 1.4116 & 20 & 74-in & 18.9$^m$\\
  \textbf{CSS0411-09} & SU & S7891$^{u}$ & 21 December 2009 & 5187.2848 & 3.6556 & 8 & 74-in & 15.3\\
  && S7893$^{u}$ & 22 December 2009 & 5188.2872 & 1.4372 & 8 & 74-in & 15.3:\\
  && S7898$^{u}$ & 24 December 2009 & 5190.2908 & 1.4061 & 8 & 74-in & 15.7:\\
  \textbf{CSS0438+00} & DN & S8022 & 28 November 2010 & 5529.4480 & 2.0579 & 30 & 40-in & 19.2$^m$\\
  && S8025 & 29 November 2010 & 5530.4410 & 3.7803 & 20 & 40-in & 19.3$^m$\\
  && S8050 & 26 December 2010 & 5557.3107 & 1.3357 & 70 & 74-in & 19.5:$^m$\\
  \textbf{CSS0449-18} & DN & S8059 & 30 January 2011 & 5592.3146 & 3.6642 & 45 & 40-in & 17.9:$^m$\\
  && S8061 & 31 January 2011 & 5593.2846 & 3.1261 & 45 & 40-in & 17.8:$^m$\\
  && S8062 & 1 February 2011 & 5594.2851 & 4.3161 & 45 & 40-in & 17.9:$^m$\\
  && S8063 & 6 February 2011 & 5599.3015 & 3.1400 & 45 & 40-in & 17.9:$^m$\\
  \textbf{SSS0501-48} & DN & S8144$^s$ & 21 January 2012 & 5948.2779 & 0.4890 & 20 & 74-in & 17.3\\
  && S8145$^s$ & 22 January 2012 & 5949.2795 & 5.5672 & 20 & 74-in & 17.2\\
  && S8147$^s$ & 23 January 2012 & 5950.2802 & 5.3910 & 20 & 74-in & 17.3\\
  \textbf{CSS0558+00} & DN & S8031 & 4 December 2010 & 5535.4867 & 1.0567 & 40 & 74-in & 19.2\\
  && S8041 & 8 December 2010 & 5539.3984 & 4.5805 & 40 & 74-in & 19.4\\
  && S8042 & 9 December 2010 & 5540.3889 & 4.6192 & 40,50,60 & 74-in & 19.6\\
  && S8044 & 12 December 2010 & 5543.4696 & 2.8842 & 60 & 74-in & 19.3\\
  && S8048 & 24 December 2010 & 5555.4143 & 2.4899 & 40 & 74-in & 18.9\\
  \textbf{CSS0902-11} & DN & S7927$^o$ & 22 March 2010 & 5278.2856 & 1.1053 & 6 & 74-in & 16.5:\\
  && S7932 & 31 March 2010 & 5287.2415 & 5.0953 & 30 & 40-in & 17.8:\\
  && S7934 & 1 April 2010 & 5288.2411 & 4.4381 & 30 & 40-in & 17.8\\
  && S8032 & 4 December 2010 & 5535.5401 & 0.9011 & 20 & 74-in & 17.6\\
  && S8039 & 7 December 2010 & 5538.5174 & 1.5656 & 10 & 74-in & 17.7:\\
  \textbf{CSS0942-19} & SU? & S8082 & 11 May 2011 & 5693.2133 & 4.1458 & 60 & 74-in & 19.6\\
  && S8085 & 12 May 2011 & 5694.2114 & 4.1342 & 60 & 74-in & 19.5\\
  && S8088 & 13 May 2011 & 5695.2133 & 3.9792 & 60 & 74-in & 19.3\\
  \textbf{CSS1052-06} & SU & S7949$^u$ & 8 April 2010 & 5295.2194 & 3.4922 & 10 & 40-in & 16.0\\
  && S7952$^u$ & 9 April 2010 & 5296.2272 & 1.9869 & 10 & 40-in & 16.3:\\
  && S7955$^u$ & 10 April 2010 & 5297.4247 & 0.6506 & 10 & 40-in & 16.4\\
  && S7964 & 13 April 2010 & 5300.3142 & 3.2947 & 20 & 40-in & 18.7\\
  \textbf{SSS1128-34} & DN & S8091 & 14 May 2011 & 5696.2570 & 4.9031 & 30 & 74-in & 18.7:\\
  && S8094$^o$ & 15 May 2011 & 5697.2270 & 3.1458 & 40 & 74-in & 16.0:\\ 
  && S8095$^o$ & 16 May 2011 & 5698.2932 & 2.5563 & 40 & 74-in & 16.0:\\
  \textbf{CSS1221-10} & DN & S8056 & 28 January 2011 & 5590.5679 & 1.4753 & 90 & 40-in & 19.3:\\
  && S8058 & 29 January 2011 & 5591.4973 & 3.0451 & 120 & 40-in & 19.3:\\
  && S8060 & 30 January 2011 & 5592.5197 & 1.8894 & 100 & 40-in & 19.3:\\
  && S8064 & 6 February 2011 & 5599.4637 & 3.9494 & 120 & 40-in & 19.4:\\
  && S8066 & 7 February 2011 & 5600.4846 & 1.2688 & 120 & 40-in & 19.7:\\
  && S8068 & 5 March 2011 & 5626.4949 & 3.6354	& 120 & 40-in& 19.6:\\
  && S8069 & 6 March 2011 & 5627.3897 & 4.5690	& 120 & 40-in& 19.4:\\
  && S8071 & 8 March 2011 & 5629.3847 & 6.4775	& 120 & 40-in& 19.4\\
  \textbf{SSS1224-41} & DN & S8162$^s$ & 28 February 2012 & 5986.5230 & 2.4560 & 120 & 40-in & 19.4$^m$\\
  && S8177$^s$ & 17 March 2012 & 6004.4454 & 5.1158 & 120 & 74-in & 19.3$^m$\\
  && S8180$^s$ & 18 March 2012 & 6005.4696 & 4.1647 & 120 & 74-in & 19.3$^m$\\
  && S8182$^s$ & 19 March 2012 & 6006.3926 & 6.0487 & 120 & 74-in & 19.3$^m$\\
  && S8185$^s$ & 20 March 2012 & 6007.4541 & 4.5465 & 120 & 74-in & 19.3$^m$\\
  \textbf{SSS1340-35} & DN & S8188$^s$ & 20 April 2012 & 6038.3205 & 2.9894 & 60 & 40-in & 18.4$^m$\\
  \textbf{CSS1417-18} & SU & S7992$^o$ & 6 July 2010 & 5384.2382 & 2.3681 & 20 & 74-in & 16.8\\
  && S7993$^o$ & 7 July 2010 & 5385.2267 & 2.5268 & 20 & 74-in & 18.8\\
  && S8080 & 10 May 2011 & 5692.3198 & 4.5998 & 120 & 74-in & 19.9\\
  && S8083 & 11 May 2011 & 5693.4119 & 3.0331 & 120 & 74-in & 20\\
  && S8089 & 13 May 2011 & 5695.4009 & 3.0666 & 120 & 74-in & 19.7\\
\hline
\multicolumn{9}{p{14cm}}{\footnotesize{\textit{Notes:} t$_{in}$: Integration time, DN: Dwarf Nova, SU: SU Ursae Majoris, P: Polar, $^s$observations were taken with the SHOC camera (as opposed to the UCT CCD), $^o$system was in outburst, $^{u}$system was in superoutburst, $^m$mean magnitude out of eclipse, : denotes an uncertain value (see Sec.~\ref{sec:observations} for details).}}\\
\end{tabular}
\end{minipage}
\end{table*}

\begin{table*}
\contcaption{}
 \centering
 \begin{minipage}{140mm}
  \begin{tabular}{lllllllll}
  \hline
  Object & Type & Run No. & Date of obs. & HJD of first obs. & length & t$_{in}$ & Tel. & $r$\\
  & & & (start of night) & (+2450000.0) & (h) & (s) & & (mag)\\
  \hline
  \textbf{CSS1556-08} & DN & S8078$^o$ & 9 May 2011 & 5691.4659 & 4.3199 & 10 & 74-in & 16.9:\\
  && S8081 & 10 May 2011 & 5692.5194 & 2.9558 & 10 & 74-in & 18.0\\
  && S8086 & 12 May 2011 & 5694.3974 & 3.7759 & 10 & 74-in & 18.4:\\
  \textbf{CSS1727+13} & DN & S7967 & 19 May 2010 & 5336.5160 & 2.3258 & 45 & 74-in & 19.0:\\
  && S7970 & 20 May 2010 & 5337.4814 & 3.4521 & 45 & 74-in & 19.7:\\
  && S7972 & 21 May 2010 & 5338.4636 & 3.0020 & 45 & 74-in & 19.7:\\
  && S7976 & 22 May 2010 & 5339.4875 & 2.1676 & 55 & 74-in & 19.7:\\
  \textbf{SSS1944-42} & P & S8084 & 11 May 2011 & 5693.5527 & 2.9464 & 10 & 74-in & 17.4\\
  && S8087 & 12 May 2011 & 5694.5753 & 2.4897 & 10 & 74-in & 17.6:\\
  && S8090 & 13 May 2011 & 5695.5555 & 3.1700 & 10 & 74-in & 17.2\\
  \textbf{SSS2003-28} & SU & S8093 & 14 May 2011 & 5696.5973 & 2.0719 & 35 & 74-in & 19.1:$^m$\\
  && S8096$^o$ & 16 May 2011 & 5698.5522 & 2.9172 & 90,100 & 74-in & 16.4:$^m$\\
  \textbf{CSS2054-19} & SU & S8005$^u$ & 31 October 2010 & 5501.2447 & 4.0508 & 45 & 74-in & 16.5:\\
  && S8008$^u$ & 1 November 2010 & 5502.2457 & 3.4636 & 45 & 74-in & 16.7\\
  && S8010$^u$ & 2 November 2010 & 5503.2729 & 2.5758 & 30 & 74-in & 17.3:\\
  \textbf{CSS2108-03} & DN & S8107$^{s}$ & 11 October 2011 & 5846.2343 & 4.5482 & 30 & 74-in & 17.7$^m$\\
  && S8109$^{s}$ & 15 October 2011 & 5850.2412 & 2.4834 & 60 & 40-in & 17.6$^m$\\
  && S8111$^{s}$ & 16 October 2011 & 5851.2308 & 3.7767 & 60 & 40-in & 17.6$^m$\\
  && S8113$^{s}$ & 19 October 2011 & 5854.2789 & 0.6000 & 60 & 40-in & 17.8$^m$\\
\hline
\multicolumn{9}{p{14cm}}{\footnotesize{\textit{Notes:} t$_{in}$: Integration time, DN: Dwarf Nova, SU: SU Ursae Majoris, P: Polar, $^s$observations were taken with the SHOC camera (as opposed to the UCT CCD), $^o$system was in outburst, $^{u}$system was in superoutburst, $^m$mean magnitude out of eclipse, : denotes an uncertain value (see Sec.~\ref{sec:observations} for details).}}\\
\end{tabular}
\end{minipage}
\end{table*}

\subsection{CSS0116+09 (CSS081220:011614+092216)}

CSS0116+09 was discovered by the CRTS on December 20, 2008 when it was observed in outburst at $\Delta V\sim$3 mag. It has a counterpart in Data Release 8 of the Sloan Digital Sky Survey (SDSS, \citet{Aihara2011}) with $u$=19.0, $g$=19.1 and $r$=19.0. \citet{Thorstensen2012} further observed it at $B-V$=0.24$\pm0.06$, $V$=18.89$\pm0.02$, $V-I$=0.64$\pm0.03$. Based on its colour, they confirmed its CRTS classification as a CV.

\begin{figure}
  \includegraphics{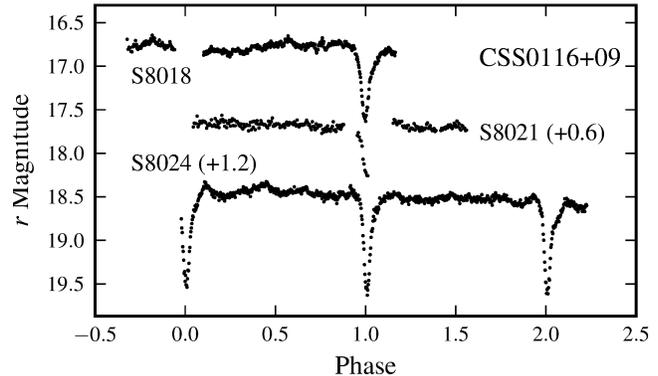}
  \caption{Light curves for CSS0116+09 taken on the decline from outburst, showing deep eclipses. The runs are offset vertically for display purposes by the magnitude indicated in parentheses. The gaps in the light curves were caused by passing cloud.}
  \label{fig:CSS0116_outburst}
\end{figure}

We caught CSS0116+09 on the decline from outburst in November 2010 and in quiescence on two further observing runs (see Tab. 1). The outburst light curves are given in Fig.~\ref{fig:CSS0116_outburst} and show CSS0116+09 to be a deeply eclipsing system. 

\begin{figure}
  \includegraphics{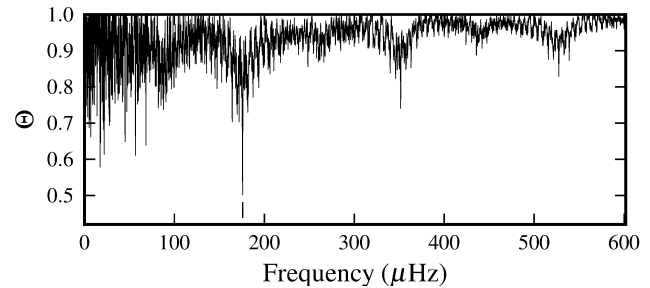}
  \caption{Phase Dispersion Minimization periodogram for combined runs S8018, S8021, S8024, S8033 and S8040 of CSS0116+09. The orbital period at 175.84$\pm$0.13$\mu$Hz is indicated by a vertical bar.}
  \label{fig:CSS0116_pdm}
\end{figure}

The PDM periodogram (see Fig. \ref{fig:CSS0116_pdm}) of these observations together with those taken in quiescence a week later (S8033 and S8040), gives an orbital period of 0.0657$\pm$0.0001$\,$d. Bootstrapping the results gives P$_{orb}$=0.06582$\pm$0.00005$\,$d, where the orbital period and uncertainty are the median and standard deviation of the distribution respectively. The ephemeris for minimum light is\footnote{E is the cycle number and the error on the last digit is quoted in parentheses, for example 0.06582($\pm$5) denotes an error of 0.00005.}
\begin{equation}\label{eq:CSS0116+09ephem}
 HJD_{min} = 2\,455\,528.3681+0\fd06582(\pm5)E\mbox{ .}
\end{equation}

\begin{figure}
  \includegraphics{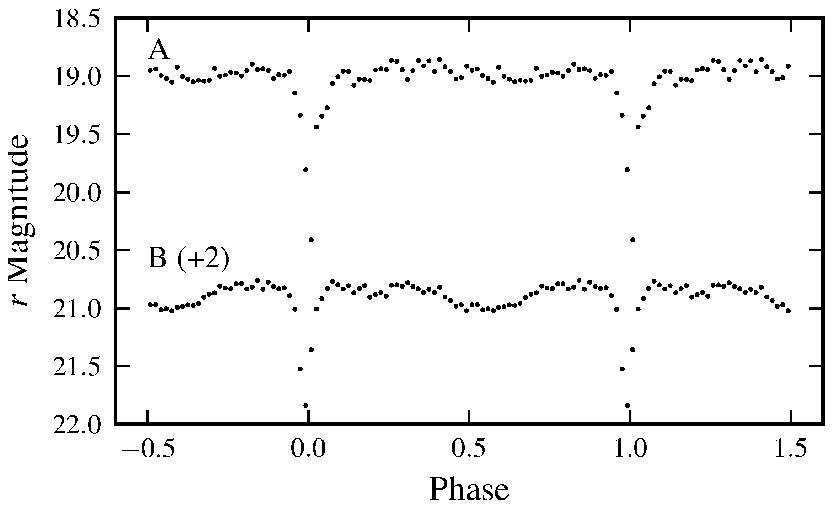}
  \caption{Average binned light curves of the quiescent runs of CSS0116+09 folded on the orbital period. Each orbit is plotted twice for display purposes. Plot A: Average of S8033 and S8040. Plot B: Average of S8098 and S8100, offset by 2 mag for display purposes.}
  \label{fig:CSS0116_quiescent}
\end{figure}

Plot A of Fig.~\ref{fig:CSS0116_quiescent} shows the average light curve of S8033 and S8040 folded on the orbital period. The average of runs S8098 and S8100 is shown in Plot B of the figure. Unfortunately these runs were too far removed in time to reduce the uncertainty in Eq. \ref{eq:CSS0116+09ephem}. From S8040 to S8098, the uncertainty on the eclipse timing amounted to more than a period and so combining these runs would have introduced a cycle ambiguity.

The eclipse depth of the light curves ranges from 0.8$\,$mag to 1.4$\,$mag. It is at its minimum in outburst, when the accretion disc contributes a larger fraction of the light. This depth indicates that CSS0116+09 has an inclination of greater than approximately 70$\degr$.

\subsection{CSS0411-09 (CSS091215:041134-090729)}

The CRTS light curve\footnote{http://nesssi.cacr.caltech.edu/catalina/AllCV.html} of CSS0411-09 has a number of high amplitude outbursts (up to 3.8$\,$mag) over the 8-year baseline. No follow-up work had been undertaken on this object prior to this paper.
 
\begin{figure}
  \includegraphics{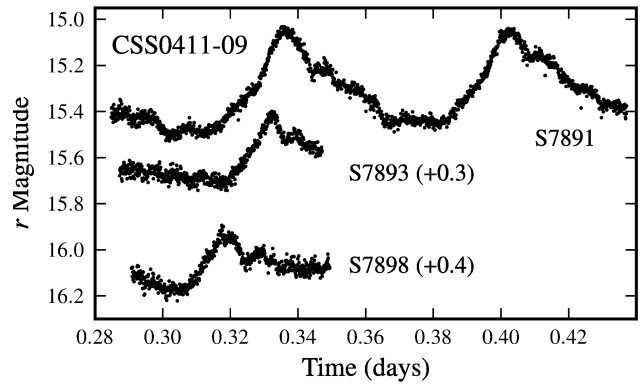}
  \caption{Observations of CSS0411-09 taken during superoutburst, showing superhumps (saw-tooth shaped profiles in the light curve produced by a precessing elliptical disc). Runs S7893 and S7898 have been displaced vertically by 0.3 and 0.4$\,$mag respectively. The time given is the fractional part of the Heliocentric Julian Date.}
  \label{fig:CSS0411}
\end{figure}

Based on our observations and the quoted CRTS quiescent magnitude of $V$=19.4, we observed the system during an outburst of more than 4$\,$mag. The light curves (Fig.~\ref{fig:CSS0411}) show superhumps with amplitudes decreasing from approximately 0.5 to 0.3$\,$mag over four days - indicating that this was a superoutburst and that CSS0411-09 is of the SU UMa class.

\begin{figure}
  \includegraphics{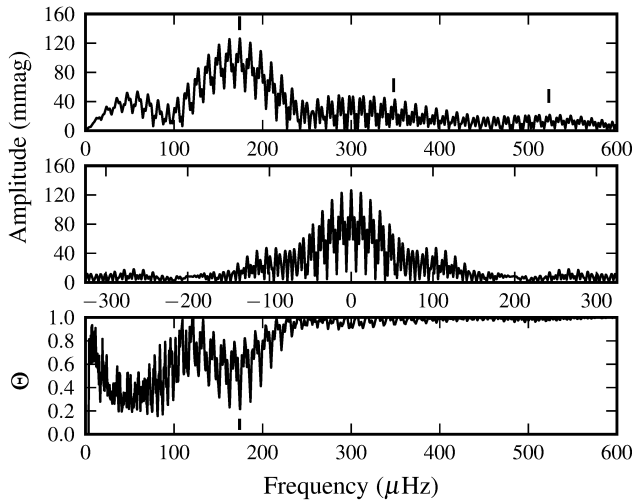}
  \caption{Upper panel: The FT of CSS0411-09 (combined runs S7891, S7893 and S7898). The superhump period at 174.40$\pm$0.02$\mu$Hz, and its first and second harmonics are marked. Middle panel: The corresponding window function shows the alias structure in the FT. Lower panel: PDM periodogram of the combined runs, with the superhump period marked.}
  \label{fig:CSS0411ft}
\end{figure}

\begin{figure}
  \includegraphics{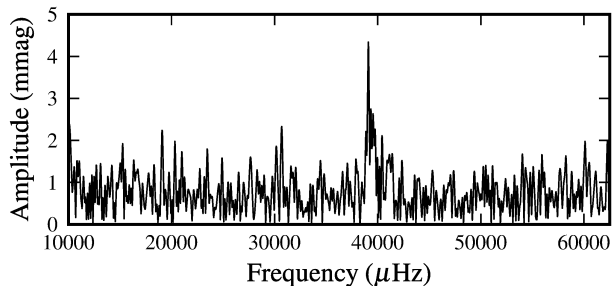}
  \caption{FT of run S7898 on CSS0411-09, showing a DNO at 39093$\,\mu$Hz (25.58$\,$s).}
  \label{fig:CSS0411ftdno}
\end{figure}

The combined FT of runs S7891, S7893 and S7898 is shown in the upper panel of Fig.~\ref{fig:CSS0411ft}. Only the mean was subtracted from each run, not the linear trend. The middle panel shows the window function, which reveals some structure in the FT and accounts for the broadness of the alias structure. Power at the first (349.18$\pm0.05\,\mu$Hz) and second harmonic (523.93$\pm$0.10$\,\mu$Hz) of the main peak at 174.40$\pm$0.02$\,\mu$Hz allows the unambiguous identification of the superhump period of 0.06637$\pm$0.00004$\,$d. The PDM periodogram (see bottom panel of Fig.~\ref{fig:CSS0411ft}) gives a consistent solution. Bootstrapping the PDM gives P$_{SH}$=0.06633$\pm$0.00001$\,$d.

Using P$_{SH}$=0.06633$\pm$0.00001$\,$d and equation 1 of \citet{Gaensicke2009} gives an approximate orbital period of 0.0645($\pm$5)$\,$d for CSS0411-09.

The FT of run S7898 (shown in Fig.~\ref{fig:CSS0411ftdno}), when CSS0411-09 started its descent towards quiescence (0.4$\,$mag fainter than the previous observation two days earlier), shows a clear peak at 25.58$\pm$0.04$\,$s with an amplitude of 4$\,$mmag. The amplitude and frequency of this rapid optical modulation is typical of dwarf nova oscillations (see \citealt{Warner2004}).
\subsection{CSS0438+00 (CSS100218:043829+004016)}

Over the course of the 6.5-year CRTS observations of CSS0438+00, three outbursts have been recorded. The first was observed at $\Delta V\sim$2.5$\,$mag above the quiescent magnitude ($V$=19.3) and the second, at $\Delta V\sim$2$\,$mag. It is within the SDSS survey area and was observed in quiescence at $u$=19.4, $g$=19.5 and $r$=19.4 mag. 

\begin{figure}
  \includegraphics{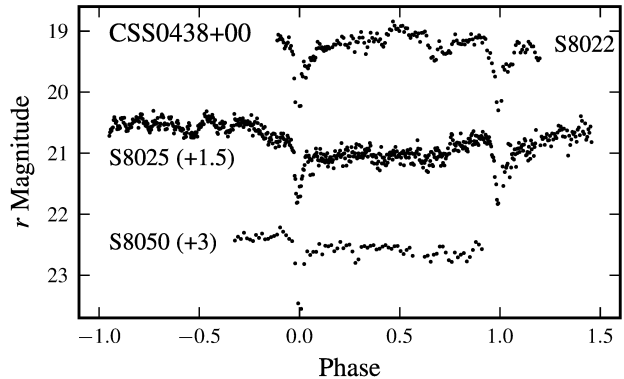}
  \caption{Light curves for CSS0438+00. Run S8050 was taken on the 74-in telescope with the UCT CCD. We observed this source under stable conditions only, as we were unable to fit a comparison star into the field-of-view and could not perform differential photometry. S8025 and S8050 are displaced vertically.}
  \label{fig:CSS0438}
\end{figure}

Fig.~\ref{fig:CSS0438} gives light curves for the three observing runs we obtained for CSS0438+00. It shows an eclipse depth of $\sim1\,$mag - indicative of a system with an inclination greater than approximately 70$\degr$. 

The magnitude of the system, when out of eclipse, varies between orbits (for example, compare S8025 before and after phase 0). This type of variation is usually attributed to changes in the mass transfer rate from the secondary. In the combined PDM periodogram of S8022 and S8025 it produced a very high amplitude peak at 0.131$\,$d. Prewhitening at this frequency gave an orbital period of 0.0657 d.

Taking the median and standard deviation of the bootstrapped sample as the orbital period and uncertainty, yields an ephemeris for minimum light of 
\begin{equation}\label{eq:CSS0438+00ephem}
 HJD_{min} = 2\,455\,529.45527+0\fd06546(\pm9)E\mbox{ .}
\end{equation}
Unfortunately S8050 could not be used to reduce the uncertainty on P$_{orb}$. Over the 27 days between S8025 and S8050, the uncertainty on the orbital period accumulated to the point where there was a cycle ambiguity.
\subsection{CSS0449-18 (CSS110114:044903-184129)}

CSS0449-18 shows frequent outbursts in its CRTS light curve, despite the fact that it is not as well sampled as the other CSS systems. The quiescent magnitude is quoted at $V$=17.7 and the outbursts were captured up to 2.8$\,$mag above quiescence.

\begin{figure}
  \includegraphics{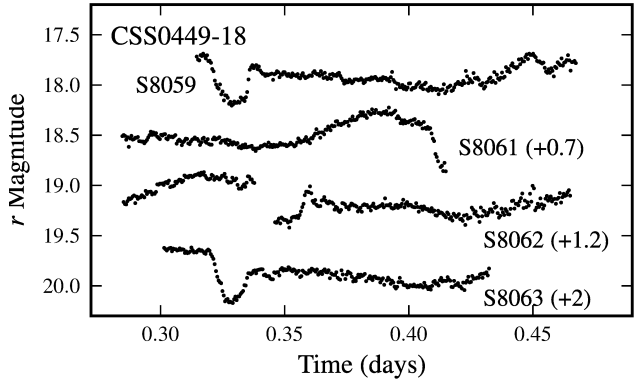}
  \caption{Photometry for CSS0449-18. S8059 is shown at the correct brightness, the remaining runs are displaced by the magnitude indicated in parentheses. The shallow, rounded eclipses are reminiscent of those of U Gem. Passing cloud caused the interruptions in the light curves.}
  \label{fig:CSS0449}
\end{figure}

Our photometry was taken at close to minimum light. The light curves show orbital humps and rounded, shallow eclipses (see Fig.~\ref{fig:CSS0449}) that vary in profile from night to night. This indicates that it is likely that only the bright spot and part of the accretion disc were eclipsed.

\begin{figure}
  \includegraphics{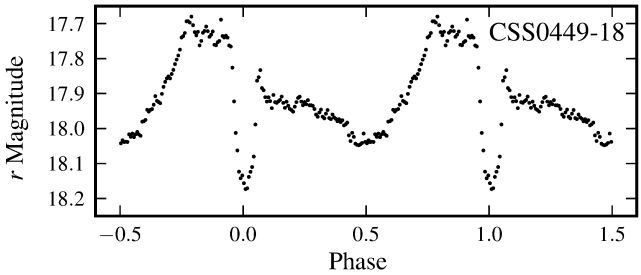}
  \caption{Average binned light curve of runs S8059, S8061, S8062 and S8063 of CSS0449-18, folded on the ephemeris given in Eq. \ref{eq:CSS0449-18ephem}. The average is plotted twice for display purposes.}
  \label{fig:CSS0449avg}
\end{figure}

From bootstrapping the PDM, we obtain an orbital period of $P_{orb}$=0.15554$\pm$0.00004$\,$d. The ephemeris for minimum light is 
\begin{equation}\label{eq:CSS0449-18ephem}
 HJD_{min} = 2\,455\,592.32760+0\fd15554(\pm4)E\mbox{ .}
\end{equation} 

The average light curve is shown in Fig.~\ref{fig:CSS0449avg}. The eclipses have a depth of $r\sim0.4\,$mag and a duration of approximately 26$\,$min. They resemble those of U Gem (see for example \citealt{Zhang1987} and \citealt{Warner1971}), consequently the inclination angle should be similar to the $\sim$72$\degr$ estimated for U Gem by \citet{Unda-Sanzana2006}.

\subsection{SSS0501-48 (SSS120112:050157-483901)}

SSS0501-48 shows a variation in its quiescent magnitude in the CRTS light curve, ranging between V$\sim$16.7 and 19$\,$mag. Two outbursts were also recorded up to V$\sim$15.3.

\begin{figure}
  \includegraphics{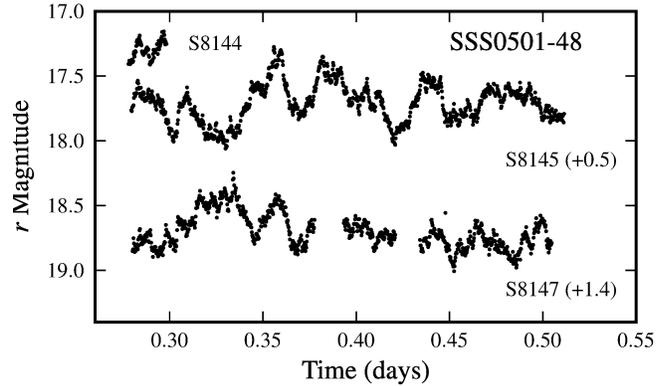}
  \caption{Light curves for SSS0501-48 dominated by flickering; no periodic modulations were found. S8145 and S8147 were 5.56$\,$h and 5.39$\,$h long respectively and are displaced vertically by 0.5 and 1.4$\,$mag for display purposes.}
  \label{fig:SSS0501}
\end{figure}

A substantial amount of flickering can be seen in our observations of SSS0501-48 (see Fig.~\ref{fig:SSS0501}), which could partly explain the scatter seen in the CRTS quiescent magnitudes. Despite the long length of the observing runs, the FTs did not show any periodic behaviour.

\subsection{CSS0558+00 (CSS100114:055843+000626)}

CSS0558+00 was discovered on January 14, 2010 during an outburst of approximately $\Delta V$=2.4$\,$mag. Two previous outbursts were also recorded in the CRTS archival data. It has a counterpart in the SDSS DR8 with $u$=20.6, $g$=20.3 and $r$=19.0. Spectroscopic follow-up observations were carried out by \citet{Thorstensen2012}, who observed it during an outburst that was not captured by the CRTS. The single spectrum they obtained was typical of a dwarf nova in outburst and showed weak, broad H$\alpha$ line in emission.

\begin{figure}
  \includegraphics{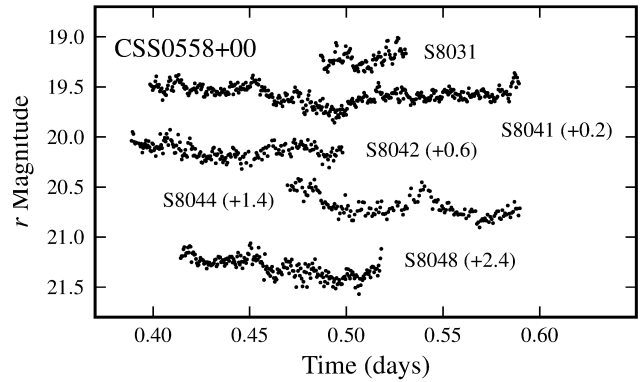}
  \caption{Light curves for CSS0558+00 taken in quiescence. The run lengths range between 1$\,$h (S8031) and 4.6$\,$h (S8041). Each run is offset vertically by the magnitude indicated in parentheses.}
  \label{fig:CSS0558}
\end{figure}

\begin{figure}
  \includegraphics{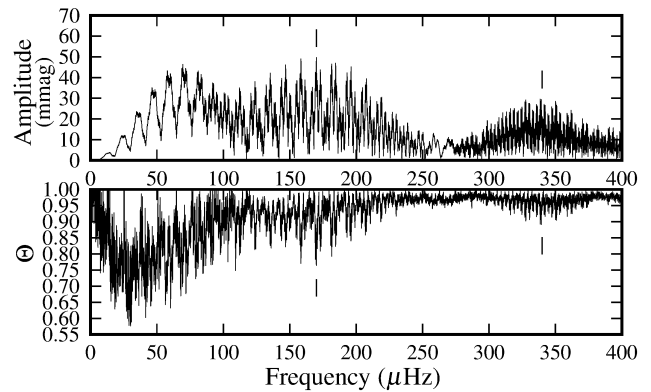}
  \caption{FT (top panel) and PDM periodogram (bottom panel) of the combined runs S8041, S8042, S8044 and S8048 of CSS0558+00. The orbital period (at 170.01$\pm$0.03$\,\mu$Hz) and its first harmonic are marked. The power at lower frequencies is caused by the finite length of the runs.}
  \label{fig:CSS0558ft}
\end{figure}

We observed this system during quiescence. Each of the runs showed flickering (see Fig.~\ref{fig:CSS0558}), which caused broad, low-amplitude peaks in the individual FTs. Figure \ref{fig:CSS0558ft} gives the FT and PDM periodogram for the four longest runs. Both contained a peak at 0.06809$\,$d and at the first harmonic. A Fisher Randomization test on the fundamental gives a false alarm probability (FAP) lying between 0 and 0.01 with 95\% confidence, that this period is not present in the data. It gives the same FAP that the period is not equal to the quoted value. These two results confirm that the 0.06809$\,$d period is real.

Bootstrapping the PDM gives an orbital period of 0.06808($\pm$1)$\,$d. Phasefolding the observations gives the ephemeris for maximum light as
\begin{equation}\label{eq:CSS0558+00ephem}
 HJD_{max} = 2\,455\,539.4522+0\fd06808(\pm1)E\mbox{ .}
\end{equation}

\subsection{CSS0902-11 (CSS090210:090210-113032)}

The quiescent magnitude for CSS0902-11, as given by the CRTS, is $V$=17.5$\,$mag. It was discovered on February 1, 2009 when it went into a 2.8$\,$mag or higher outburst.

\citet{Thorstensen2012} took time series spectroscopy of this CV, from which they estimated it to be at a distance of 1100(+350,-260) pc and have a secondary star of spectral type K7$\pm1$ and magnitude V$\sim$18.5$\,$. Additionally, based on the small velocity amplitude of the secondary (100$\pm6\,$km$\,$s$^{-1}$), they predicted that the system would not be eclipsing. This is confirmed by our photometry.

\begin{figure}
  \includegraphics{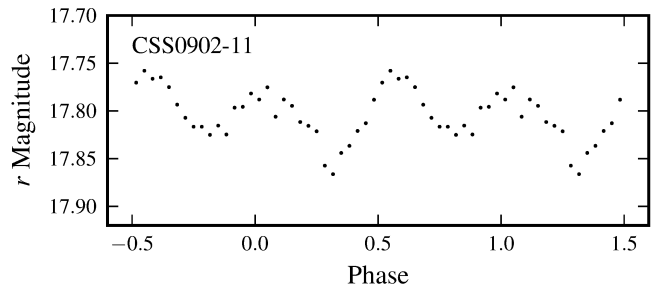}
  \caption{Average binned light curve of the two longest runs on CSS0902-11, folded on the 6.62$\pm0.01\,$h modulation detected by \citet{Thorstensen2012}. The data-length of the two runs, S7932 and S7934, were 5.1$\,$h and 4.4$\,$h respectively.}
  \label{fig:CSS0902}
\end{figure}

Their absorption spectra showed a clear 6.62$\pm0.01\,$h modulation. Only two of our runs approached this length - at 5.1 and 4.4$\,$h respectively. The combined FT gives a period of 3.3$\pm0.1\,$h, half that of \citeauthor{Thorstensen2012}'s value. Phase-folding our two longest runs (S7932 and S7934) on the 6.62$\,$h period (Fig.~\ref{fig:CSS0902}) shows that CSS0902-11 is a double hump system. Only one orbital hump was covered entirely in each of our two runs, which produced significant power in the FT at half the orbital period. In conclusion, our photometry is consistent with the 6.62$\,$h orbital period detected by \citet{Thorstensen2012}.

\subsection{CSS0942-19 (CSS090117:094252-193652)}

The CRTS light curve for CSS0942-19 is sparsely sampled, but shows two outbursts with amplitudes of approximately 4$\,$mag. Judging from the large amplitudes, these may have been superoutbursts. This system was also caught in outburst by the Palomar Quest Sky Survey \citep{Djorgovski2008}.

\begin{figure}
  \includegraphics{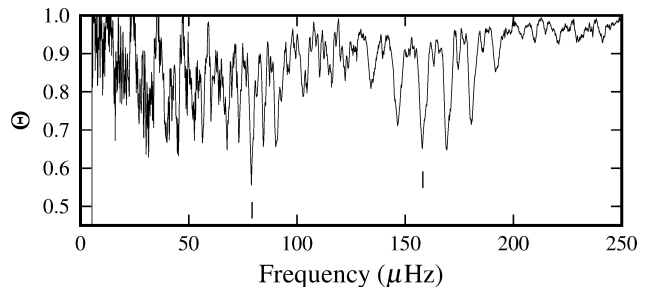}
  \caption{PDM periodogram of runs S8082, S8085 and S8088 on CSS0942-19. The peaks at the fundamental and first harmonic are marked.}
  \label{fig:CSS0942pdm}
\end{figure}

\begin{figure}
  \includegraphics{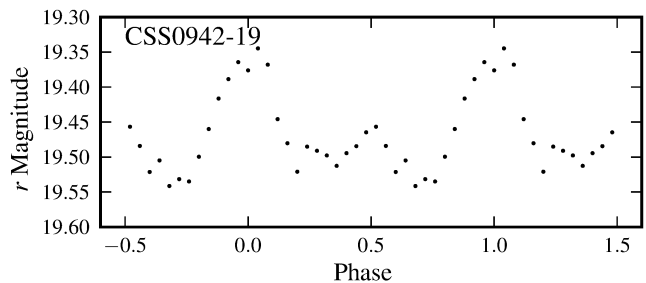}
  \caption{The average binned light curve of CSS0942-19, folded on the 0.147$\pm$0.001$\,$d photometric period.}
  \label{fig:CSS0942}
\end{figure}

We observed CSS0942-19 in quiescence in May 2011 at $V\sim19.5$ (see Tab. 1) four months after the last recorded outburst by CRTS. The PDM periodogram of the combined runs on this object (see Fig. \ref{fig:CSS0942pdm}), shows strong peaks at the orbital frequency (78.7$\pm$0.7$\,\mu$Hz) and at its first harmonic. Bootstrapping the PDM, we obtain an orbital period and ephemeris for maximum light of 
\begin{equation}\label{eq:CSS0942ephem}
 HJD_{max} = 2\,455\,693.2442+0\fd147(\pm1)E\mbox{ .}
\end{equation}
The average light curve of CSS0942-19, folded on this ephemeris, is shown in Fig.~\ref{fig:CSS0942}.

\subsection{CSS1052-06 (CSS100408:105215-064326)}

Since CSS1052-06 was detected by the CRTS on April 8, 2010, numerous outbursts of approximately $\Delta V$=3$\,$mag have been captured. No follow-up observations have been performed on it prior to this paper.

\begin{figure}
  \includegraphics{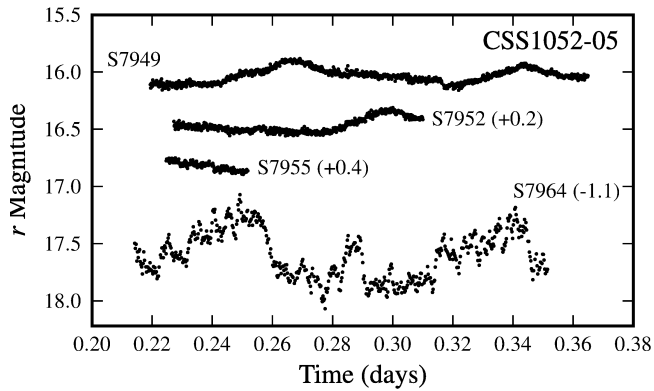}
  \caption{Light curves for CSS1052-06. S7949, S7952 and S7955 were taken during superoutburst and show superhumps. S7964 was taken in quiescence. For display purposes S7952, S7955 and S7964 have been displaced vertically by 0.2, 0.4 and -1.1$\,$mag respectively.}
  \label{fig:CSS1052}
\end{figure}

During our first three observing runs (S7949, S7952 and S7955), CSS1052-06 was in superoutburst and superhumps were present (see Fig.~\ref{fig:CSS1052}). Three days later, during S7964, the system had returned to quiescence.

\begin{figure}
  \includegraphics{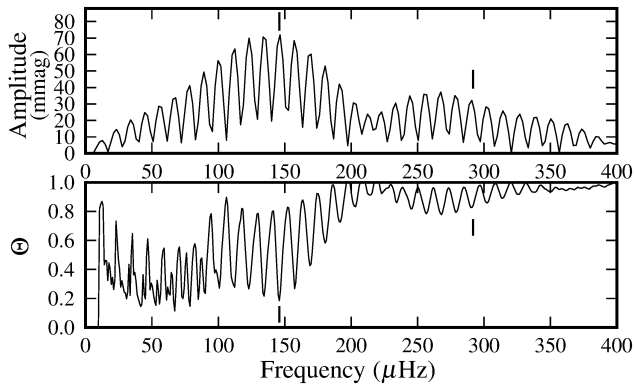}
  \caption{FT (top panel) and PDM periodogram (bottom panel) of S7949 and S7952 for CSS1052-06. Only the mean was subtracted from each run; they were not linearly detrended. The superhump period (0.07938$\,$d) and its first harmonic are marked.}
  \label{fig:CSS1052ft}
\end{figure}

The FT and the PDM periodogram of the two long runs in superoutburst are given in Fig.~\ref{fig:CSS1052ft}. The highest (or lowest in the case of the PDM) peaks coincide. Bootstrapping the PDM gives a superhump period of 0.07938($\pm$3)$\,$d. A Fisher randomisation test gives a false alarm probability, that the quoted period is different to the true period, of between 0 and 0.01 with a 95\% confidence level. 

Using the relation between the superhump and orbital periods in equation 1 of \citet{Gaensicke2009}, the approximate orbital period for this system is 0.0765$\pm0.0005\,$d. The FT of the quiescent run S7964 did not show a peak at this period. This could be attributed to the large flickering in the light curve and the run length of less than two orbits.
 
\subsection{SSS1128-34 (SSS110327:112815-344807)}

\begin{figure}
  \includegraphics{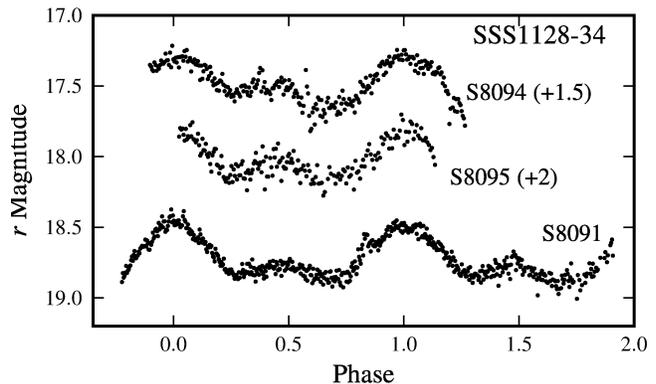}
  \caption{Light curves for SSS1128-34 phased on the orbital period (0.0985$\,$d). S8091 was taken during quiescence, whereas S8094 and S8095 were taken in outburst. The latter two runs were displaced vertically for display.}
  \label{fig:SSS1128}
\end{figure}

The CRTS have detected multiple outbursts of SSS1128-34 since it was first observed in 2006. Their light curve gives an average quiescent magnitude of $V\sim$19, although the individual observations range from $V\sim$18 to $V\sim$19.8. This range is produced by the orbital variation, as shown in our photometry in Fig.~\ref{fig:SSS1128}.

Each of the light curves show a clear double hump structure. The orbital hump of amplitude $\sim$0.4$\,$mag at phase 1 is produced by the bright spot orbiting into and out of our line of sight. The smaller hump at phase 0.5 is probably produced by the bright spot shining through a semi-transparent disc. 

During our initial run (S8091), SSS1128-34 was in quiescence. The following day the system was 1.7 mag brighter and S8094 and S8095 were taken in outburst. 

\begin{figure}
  \includegraphics{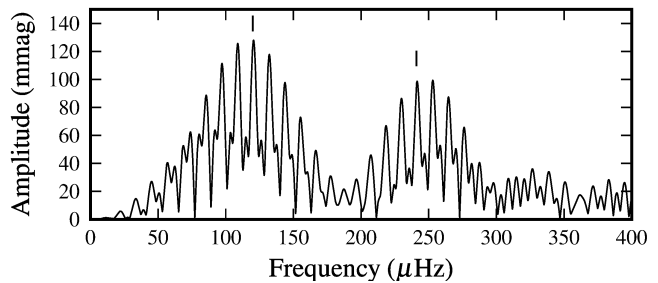}
  \caption{FT of the combined runs S8091, S8094 and S8095 of SSS1128-34. The orbital period and its first harmonic are marked.}
  \label{fig:SSS1128ft}
\end{figure}

In the FT of the combined runs, the orbital period (0.096($\pm$1)$\,$d) was chosen from two possible aliases of approximately equal height, based on the power of their respective first harmonics (see Fig.~\ref{fig:SSS1128ft}). The double hump structure produces large amounts of power at the first harmonic, so the ambiguity in the fundamental is removed because only one of the two likely orbital periods has a strong first harmonic. Only the mean, not the linear trend, was subtracted from these runs before calculating the FT, as removing the linear trend would greatly affect the shape of S8094 and S8095. 

Using the Metropolis-Hastings algorithm to sample the period probability distribution (based on sine-curve fitting of the light curve), we get P$_{orb}$=0.0985$\pm$0.0001$\,$d. The period from the FT was used as the initial guess and the median and standard deviation are quoted as the period and uncertainty. The ephemeris for maximum light for these observations is
\begin{equation}\label{eq:SSS1128ephem}
  HJD_{max}=2\,455\,696.2785\pm0.0985(\pm1)E\mbox{ ,}
\end{equation}
which places SSS1128-34 in the period gap.
\subsection{CSS1221-10 (CSS080324:122100-102735)}

The CRTS detected CSS1221-10 when it went into an outburst of magnitude $\Delta V\sim$3.5 on March 24, 2008. It does not have a counterpart in the SDSS.

\begin{figure}
  \includegraphics{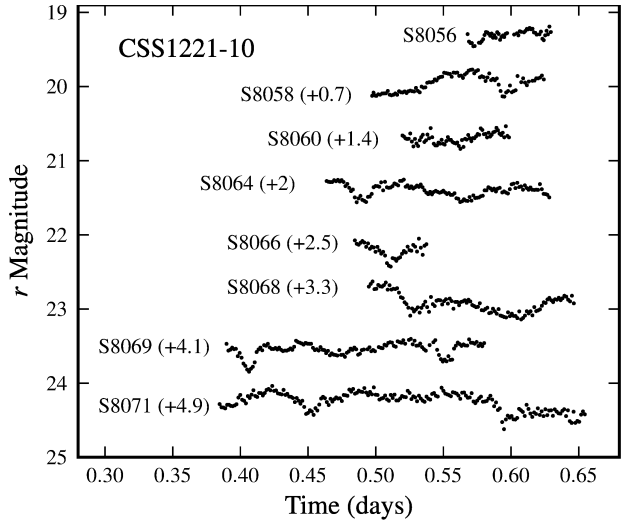}
  \caption{Light curves for CSS1221-10, showing grazing eclipses. The runs range in length from 1.3$\,$h (S8066) to 6.5$\,$h (S8071). Each run is shifted vertically by the magnitude indicated in parentheses.}
  \label{fig:CSS1221}
\end{figure}

As seen in Fig.~\ref{fig:CSS1221}, CSS1221-10 has shallow eclipses with varying profiles reminiscent of those seen in CSS0449-18 (Fig.~\ref{fig:CSS0449}). This indicates that the system has a low inclination and is showing grazing eclipses.

\begin{figure}
  \includegraphics{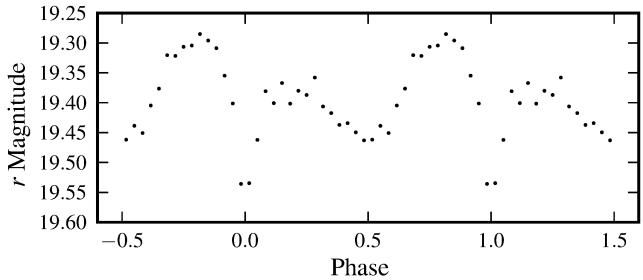}
  \caption{Averaged and binned light curve for CSS1221-10, folded on the 0.14615$\,$d orbital period. Two orbits are shown for display purposes.}
  \label{fig:CSS1221_average}
\end{figure}

The first harmonic of the orbital period was clear in the PDM periodogram of the combined runs S8056, S8058, S8064, S8068, S8069 and S8071. The ephemeris for minimum light is 
\begin{equation}\label{eq:CSS1221-10ephem}
 HJD_{min} = 2\,455\,590.5743 + 0\fd14615(\pm1)E\mbox{ ,}
\end{equation}
where the uncertainty on the orbital period was obtained by bootstrapping the results. Fig.~\ref{fig:CSS1221_average} displays the folded average light curve of the observations. The eclipse depth is approximately 0.2$\,$mag and the system shows an orbital hump.

\subsection{SSS1224-41 (SSS120215:122443-410158)}

The CRTS quiescent value for SSS1224-41 is $V$=19.1, but the light curve shows scatter between 17 and 20.6. This is due to outbursts and, as we see in our photometry, eclipses.

\begin{figure}
  \includegraphics{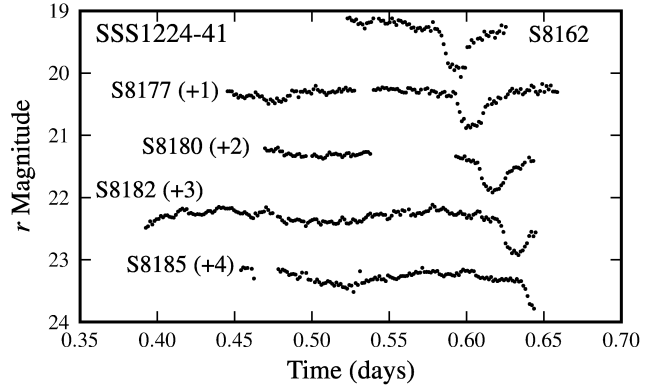}
  \caption{Light curves of SSS1224-41 showing $\sim$0.6$\,$mag eclipses. The gaps in the light curves were caused by passing cloud. The indicated vertical offsets are for display purposes.}
  \label{fig:SSS1224}
\end{figure}

We obtained five runs on SSS1224-41 over February and March 2012. The corresponding light curves, with their $\sim$0.6$\,$mag eclipses, are shown in Fig.~\ref{fig:SSS1224}. As it turns out, run S8182 was 142s short of a complete orbital cycle.

The PDM periodogram showed peaks at the orbital period (0.2537$\,$d) and its first harmonic. Bootstrapping the results to refine the orbital period and determine the uncertainty, as well as phase-folding the observations, gave an ephemeris for minimum light of
\begin{equation}\label{eq:SSS1224-41}
 HJD_{min} = 2\,455\,986.5934 + 0\fd25367(\pm3)E\mbox{ .}
\end{equation}
\subsection{SSS1340-35 (SSS120402:134015-350512)}

SSS1224-41 has a quiescent magnitude of $V\sim$18.4$\,$mag, but has a number of points that were taken at V$\leq$20. As shown by our photometry, these points were taken during eclipse.

\begin{figure}
  \includegraphics{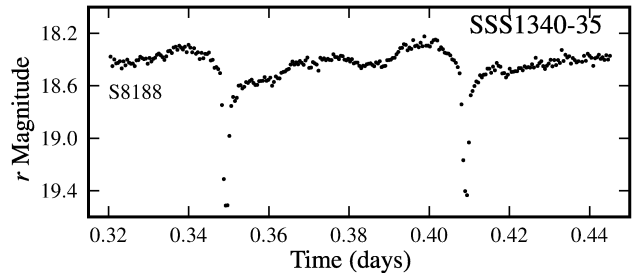}
  \caption{Light curve of SSS1340-35. The eclipse depth is $r\sim$0.9, indicating a system inclination of greater than approximately 70$\degr$.}
  \label{fig:SSS1340}
\end{figure}

We obtained one run on SSS1340-35, which contained two eclipses of depth $\sim$0.9$\,$mag (see Fig.~\ref{fig:SSS1340}). The corresponding PDM periodogram gives an orbital period of 0.059($\pm$1)$\,$d, where the uncertainty was obtained by bootstrapping. The ephemeris for minimum light is 
\begin{equation}\label{eq:SSS1340-35ephem}
 HJD_{min} = 2\,456\,038.3492 + 0\fd059(\pm1)E\mbox{ .}
\end{equation}

\subsection{CSS1417-18 (CSS080425:141712-180328)}

The CRTS detected one superoutburst of CSS1417-18 at $V\approx$15 - approximately 5 magnitudes higher than the SDSS DR8 quiescent magnitude of $r$=20.3. We obtained two sets of observations of this object, one in outburst at $r$=16.8 and one in quiescence at $r$=20. This outburst was not captured by the CRTS.

\begin{figure}
  \includegraphics{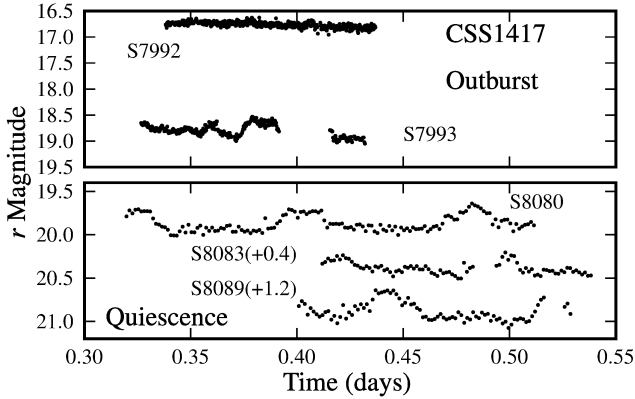}
  \caption{Light curves for CSS1417-18. The top panel contains those taken in outburst and the bottom panel contains those taken in quiescence. S8083 and S8089 have been displaced vertically by 0.4 and 1.2$\,$mag respectively.}
  \label{fig:CSS1417}
\end{figure}

All the observations are given in Fig.~\ref{fig:CSS1417}. Consider first the quiescent light curves shown in the bottom panel. They show a clear orbital hump profile with a $\sim$0.2$\,$mag amplitude. Using the highest amplitude peak in the FT as the initial guess for sine-curve fitting, the MCMC method produced a distribution of periods with median value 0.0845$\,$d and standard deviation 0.0001$\,$d, which we quote as the orbital period and uncertainty. The ephemeris for maximum light is
\begin{equation}\label{eq:CSS1417ephem}
 HJD_{max} = 2\,455\,692.3254 + 0\fd0845(\pm1)E\mbox{ .}
\end{equation}

Now consider the two runs taken in outburst. S7992 did not show a modulation near the orbital period because, at $r$=16.8, the high disc luminosity overwhelmed all other variations. The gap in S7993, caused by passing cloud, was too large to pick up an orbital variation in this run.
 
In the one day separating S7992 and S7993, CSS1417-18 declined in magnitude by $r$=2$\,$mag. In dwarf novae (DNe) the decline rate from outburst is related to the orbital period via equation 3.5 in \citet{Warner1995}:
\begin{equation}\label{eq:declinerate}
  \tau_d=0.53P^{0.84}_{orb}(h)\mbox{ d}\,\mbox{mag}^{-1}\mbox{ .}
\end{equation}

Using the 0.0845$\,$d orbital period obtained from the quiescent runs, we would expect a decline rate of $\sim$0.96$\,$mag$\,$d$^{-1}$, instead of the observed 2$\,$mag$\,$d$^{-1}$. This value falls below the scatter on the plot of log($\tau_d$) versus log(P$_{orb}$) for DNe (see Figure 3.11 of \citet{Warner1995}). Further observations of this system in decline are needed.
\subsection{CSS1556-08 (CSS090321:155631-080440)}

CSS1556-08 is listed as $V$=18.4 by the CRTS. On January 31, 2011, \citet{Thorstensen2012} took two spectra of it showing broad H$\alpha$ and H$\beta$ lines in emission. In March 2012 it went into superoutburst at V$\approx$15.3$\,$mag, which was observed by \citet{Ohshima2012a}, \citet{Ohshima2012b} and \citet{Kato2013} (the system is refered to as OT J155631.0-080440). \citet{Ohshima2012a} gave an initial value for the superhump period as 0.1 d, which was then refined to 0.08933$\pm$0.00006$\,$d by \citet{Ohshima2012b}. \citet{Kato2013} determined the superhump period to be 0.089309$\pm$0.000053$\,$d. The average profile phased on this period is shown in figure 70 of \citet{Kato2013}.   

\begin{figure}
  \includegraphics{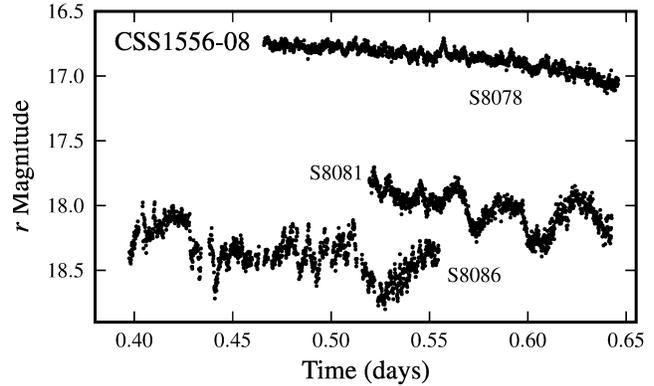}
  \caption{Light curves for CSS1556-08. Run S8078 is in outburst.}
    \label{fig:CSS1556}
\end{figure}

Our photometry of CSS1556-8 includes one run in outburst (S8078) and two in quiescence (S8086 and S8081) - it is shown in Fig.~\ref{fig:CSS1556}.

Using equation 1 from \citet{Gaensicke2009} and the 0.089309$\pm$0.000053$\,$d superhump period from \citet{Kato2013}, gives an estimate for the orbital period as 0.0856$\pm$0.0006$\,$d. Although this period is present in the FTs of S8078 and of S8081 and S8086 combined, it is not the highest power alias and does not have any high power harmonics.
\subsection{CSS1727+13 (CSS090929:172734+130513)}

The CRTS quiescent magnitude for CSS1727+13 is listed as $V$=19.5, but is probably lower because many of the observations only yielded upper limits - most of them at $V\sim$20.3.

\begin{figure}
  \includegraphics{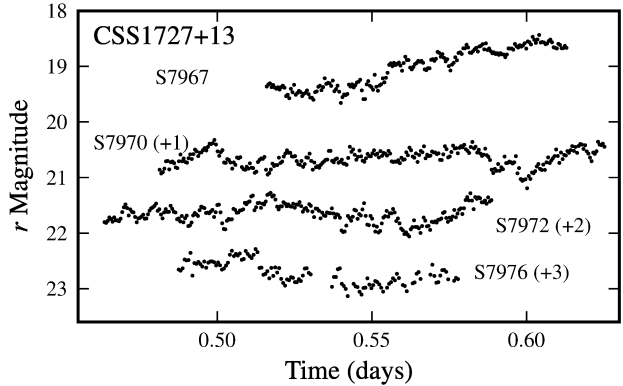}
  \caption{Light curves for CSS1727+13. No coherent modulations were indicated by the FT. Vertical offsets are indicated in parentheses.}
  \label{fig:CSS1727}
\end{figure}

\citet{Thorstensen2012} observed this object spectroscopically. Based on the FWHM of the H$\alpha$ line (1200$\,$km$\,$s$^{-1}$), they concluded that CSS1727+13 has an intermediate inclination. Our light curves (Fig.~\ref{fig:CSS1727}) show flickering, but the FT and PDM of the combined runs only show low amplitude, broad peaks.

\subsection{SSS1944-42 (SSS100805:194428-420209)}

The $\sim$7-year long CRTS light curve for SSS1944-42 shows two low states and a high state lasting from August 2009 to June 2011; no outbursts have been captured. This behaviour is typical of a magnetic system and the absence of outbursts indicates that it may be a polar. Our observations were taken during the high state.

\begin{figure}
  \includegraphics{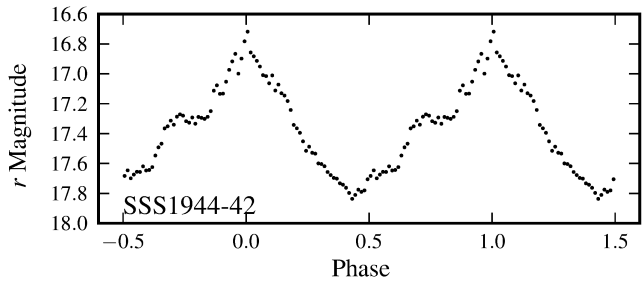}
  \caption{Average, binned light curve of SSS1944 folded on the ephemeris in Eq. \ref{eq:SSS1944ephem}. Two orbits are shown for display purposes.}
  \label{fig:SSS1944}
\end{figure}

Only the mean of each of the individual runs was subtracted, not the linear trend. The PDM periodogram of the combined runs has a peak at the orbital period (0.063855$\,$d). Bootstrapping yields P$_{orb}$=0.06385$\pm$0.00002$\,$d. The ephemeris for maximum light for this system is
\begin{equation}\label{eq:SSS1944ephem}
 HJD_{max}=2\,455\,693.6148+0\fd06385(\pm2)E \mbox{ .}
\end{equation}
Fig.~\ref{fig:SSS1944} shows the average light curve folded on this ephemeris.

\subsection{SSS2003-28 (SSS100615:200331-284941)}

Over the course of the CRTS observations (from April 2005 to September 2012), only one outburst has been recorded for this object. It occurred on the 15th of June 2010, when the system rose to $V$=15.4 from the listed quiescent value of $V$=18.8.

\begin{figure}
  \includegraphics{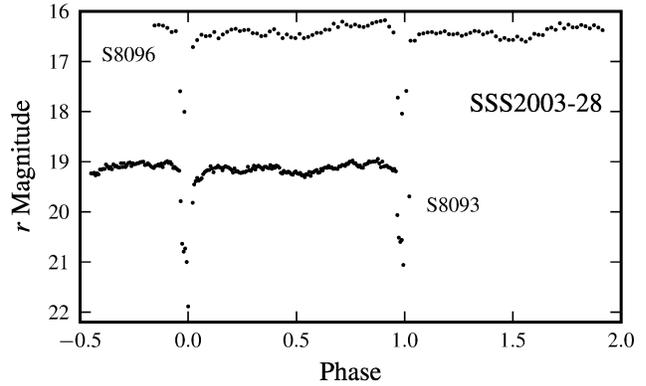}
  \caption{Lightcurves of SSS2003-28 folded on the orbital period. S8096 was taken during outburst. The average eclipse depth is approximately 1.9 mag.}
  \label{fig:SSS2003}
\end{figure}

On the 16th of May 2011, we caught this system in an outburst that was not captured by the CRTS. This run (S8096), as well as the quiescent run taken two days previously, are both displayed in Fig.~\ref{fig:SSS2003}. The photometry shows eclipses of depth $r\sim$1.9, indicating an inclination of greater than approximately 70$\degr$.

The ephemeris for the time of mid-eclipse for these observations is
\begin{equation}\label{eq:SSS2003ephem}
 HJD_{min}=2\,455\,696.6236+0\fd05871(\pm4)E \mbox{ ,}
\end{equation}
where the orbital period and uncertainty were obtained by bootstrapping the PDM.
\subsection{CSS2054-19 (CSS090829:205408-194027)}

\begin{figure}
  \includegraphics{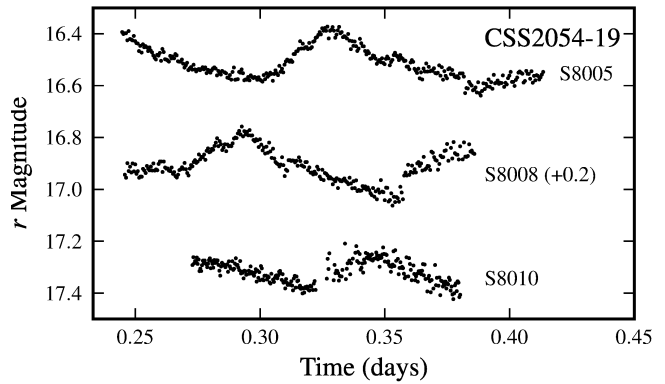}
  \caption{Light curves for CSS2054-19 taken during outburst, showing superhumps of amplitude $\sim$0.2$\,$mag. S8005 and S8010 are displayed at the correct brightness; S8008 has been shifted by 0.2$\,$mag.}
  \label{fig:CSS2054}
\end{figure}

Since the detection of CSS2054-19 on the 29th of August 2009, numerous outbursts have been captured by the CRTS, the highest amplitude being $\Delta V$=3.5 above the quoted quiescent value. The large amplitudes indicated that this may be an SU UMa system. Our observations confirm this. The photometry (Fig.~\ref{fig:CSS2054}) shows superhumps of amplitude $\sim$0.2$\,$mag, indicating that CSS2054-19 was in superoutburst. 

\begin{figure}
  \includegraphics{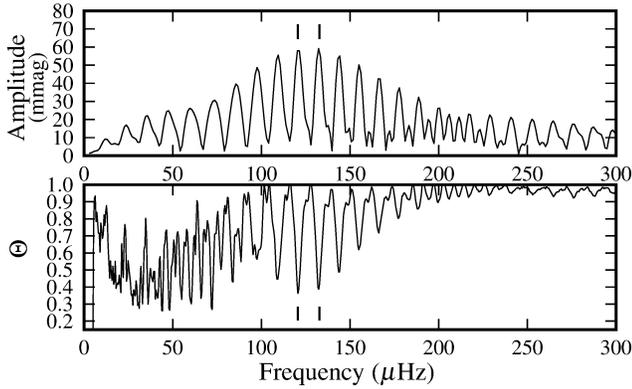}
  \caption{FT (top panel) and PDM periodogram (bottom panel) of the combined runs S8005, S8008 and S8010 of CSS2054-19. The vertical bars mark the two possible aliases for the superhump period, at 0.096$\pm$0.001$\,$d and 0.0872$\pm$0.0009$\,$d.}
  \label{fig:CSS2054ftandpdm}
\end{figure}

The FT of the combined runs on this object gave two possible aliases for the superhump period, namely 0.096$\pm$0.001$\,$d and 0.0872$\pm$0.0009$\,$d, where the former was marginally higher. The corresponding PDM periodogram gave the same results (see Fig. \ref{fig:CSS2054ftandpdm}). Bootstrapping the PDM produced a peak at both aliases. The first peak, at 0.09598$\pm$0.00008$\,$d, accounted for 88.5\% of the data points, while the second peak (at 0.08752$\pm$0.00008$\,$d) accounted for the remainder. Based on these results, we have a preference for the 0.09598$\pm$0.00008$\,$d period, but can't exclude the one day alias.

Using P$_{SH}$=0.09598$\pm$0.00008$\,$d, equation 1 of \citet{Gaensicke2009} estimates an orbital period of 0.0917$\pm$0.0006$\,$d. This estimate would place CSS2054-19 within the period gap.
\subsection{CSS2108-03 (CSS110513:210846-035031)}

\begin{figure}
  \includegraphics{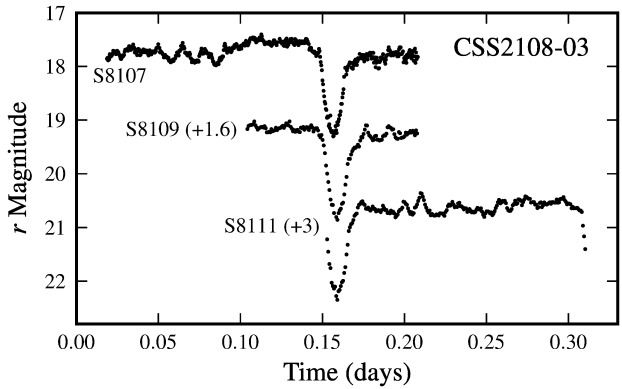}
  \caption{Light curves for CSS2108-03 folded on the ephemeris given in Eq. \ref{eq:CSS2108ephem}. The shortest run (S8113) is not shown. S8109 and S8111 have been shifted by 1.6 and 3$\,$mag for display purposes.}
  \label{fig:CSS2108}
\end{figure}

The CRTS light curve for CSS2108-03 gives the quiescent magnitude as $V$=18 and its highest amplitude outburst at $V$=14.9. Its counterpart in the SDSS DR8 was observed in quiescence at $u$=18.7, $g$=18.6 and $r$=18 \citep{Aihara2011}. The CRTS light curve for this object gives strong indications that it is an eclipsing system, as it has a number of points that are found more than a magnitude below quiescence. The faintest point is an upper-limit for detection at $V$=20.7. Photometry on CSS2108-03 confirms this, showing deep eclipses and large amplitude flickering (see Fig. ~\ref{fig:CSS2108}). 

\begin{figure}
  \includegraphics{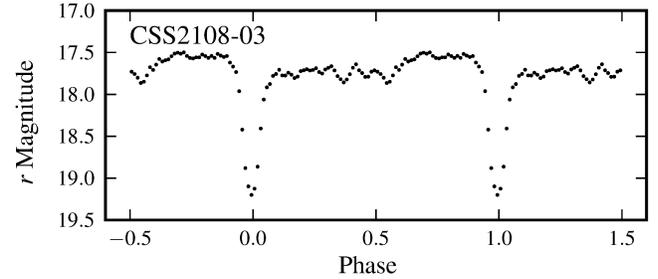}
  \caption{Average, binned light curve for CSS2108-03, folded on the ephemeris given in Eq. \ref{eq:CSS2108ephem}. The eclipse depth is $\sim$1.5$\,$mag.}
  \label{fig:CSS2108_average}
\end{figure}

\citet{Kato2013vsnet} observed this system photometrically and determined an orbital period of 0.156926798($\pm$9)$\,$d. Independently we determined an orbital period of 0.15699($\pm$5)$\,$d by bootstrapping the PDM of runs S8107, S8109 and S8111 combined. The ephemeris of minimum light for our observations is
\begin{equation}\label{eq:CSS2108ephem}
 HJD_{min}=2\,455\,846.3725+0\fd15699(\pm5)E \mbox{ .}
\end{equation}
The average of the runs folded on this ephemeris is presented in Fig.~\ref{fig:CSS2108_average}. It has an eclipse depth of $\sim$1.5$\,$mag.

\section{Discussion and Conclusions}\label{sec:conclusions}

\begin{table*}\label{tbl:Summary}
 \centering
 \begin{minipage}{140mm}
  \caption{Summary of results}
  \begin{tabular}{llllll}
  \hline
  Object & Type & P$_{orb}$ & P$_{SH}$ & $r$ & Remarks\\
  && (d) & (d) &&\\
  \hline
  CSS0116+09 & DN & 0.06582($\pm$5) & - & 18.9$^{m,q}$, 17.1$^{m,o}$ & Eclipsing, depth$\sim$1.4 mag\\
  CSS0411-09 & SU & - & 0.06633($\pm$1) & 15.4$^o$& Superhump\\ 
  CSS0438+00 & DN & 0.06546($\pm$9) & - & 19.3$^{m,q}$ & Eclipsing, depth$\sim$1 mag\\
  CSS0449-18 & DN & 0.15554($\pm$4) & - & 17.9$^{m,q}$ & Eclipsing, depth$\sim$0.4 mag\\
  SSS0501-48 & DN & - & - & 17.3$^q$ &\\
  CSS0558+00 & DN & 0.06808($\pm$1) & - & 19.3$^q$ &\\
  CSS0902-11 & DN & 0.2758($\pm$4)$^*$ & - & 17.7$^q$, 16.5$^o$ &\\
  CSS0942-19 & SU? & 0.147($\pm$1) & - & 19.5$^q$ &\\
  CSS1052-06 & SU & - & 0.07938($\pm$3) &16.2$^o$, 18.7$^q$& Superhump\\
  SSS1128-34 & DN & 0.0985($\pm$1) & - & 18.7$^q$, 16$^o$&\\
  CSS1221-10 & DN & 0.14615($\pm$1) & - & 19.4$^{q,m}$ & Eclipsing, depth$\sim$0.2 mag\\
  SSS1224-41 & DN & 0.25367($\pm$3) & - & 19.3$^{q,m}$ & Eclipsing, depth$\sim$0.6 mag\\
  SSS1340-35 & DN & 0.059($\pm$1) & - & 18.4$^{q,m}$ & Eclipsing, depth$\sim$0.9 mag\\
  CSS1417-18 & SU & 0.0845($\pm$1) & - & 17.8$^o$, 19.9$^q$ & Fast decline from outburst\\
  CSS1556-08 & SU & - & 0.089309($\pm$53)$^{**}$ & 16.9$^o$, 18.2$^q$ & \\
  CSS1727+13 & DN & - & - & 19.5$^q$ &\\
  SSS1944-42 & P & 0.06385($\pm$2) & - & 17.4$^q$&\\
  SSS2003-28 & SU & 0.05871($\pm$4) & - & 19.1$^{q,m}$, 16.4$^{o,m}$& Eclipsing, depth$\sim$1.9 mag\\
  CSS2054-19 & SU & - & 0.09598($\pm$8)$^{\diamond}$ & 16.8$^o$ & Superhump\\ 
  CSS2108-03 & DN & 0.15699($\pm$5)$^{\diamond\diamond}$ & - & 17.7$^{q,m}$ & Eclipsing, depth$\sim$1.5 mag\\
  \hline
  \multicolumn{6}{p{14cm}}{\footnotesize{\textit{Notes:} Uncertainties on the last decimal are given in parentheses. DN: Dwarf Nova, SU: SU Ursae Majoris, P: Polar, $^m$mean magnitude out of eclipse, $^o$outburst magnitude, $^q$quiescent magnitude, $^*$period determined by \citet{Thorstensen2012}}, $^{**}$period from \citet{Kato2013}, $^{\diamond}$one day alias is at 0.08752($\pm$8)$\,$d, $^{\diamond\diamond}$\citet{Kato2013vsnet} independently found P$_{orb}=$0.156926798($\pm$9)$\,$d}\\
\end{tabular}
\end{minipage}
\end{table*}

We observed 20 CVs identified by the CRTS with the aim of classifying them, determining orbital periods and selecting targets for further observations with large telescopes. Of these 20 systems, only 6 have been observed prior to this work. Four systems were confirmed to be CVs by means of spectra and standardized photometry and a superhump period was determined for one of these 
(CSS1556-08, \citealt{Kato2013}). CSS0902-11 has been followed-up in detail with time-series spectroscopy \citep{Thorstensen2012} and CSS2108-03 was observed photometrically by \citet{Kato2013vsnet}. 

The results are summarised in Tab. 2. We determined 12 new orbital periods and independently discovered periods for a further two CVs, namely CSS0902-11 and CSS2108-03. For three of the systems (CSS0411-09, CSS1052-06 and CSS2054-19), we determined superhump periods. CSS1556-08 had a pre-determined superhump period (\citealt{Ohshima2012b} and \citealt{Kato2013}), but we could not determine an orbital period from our photometry. The remaining 2 CVs did not show any periodic modulations.

Most of the CVs were DNe systems. This is as we should expect, as the CRTS identifies transients based on their variability. There was also a polar in the sample (SSS1944-42), that was picked up by the CRTS because it showed high and low states that differed by more than a magnitude.

The orbital periods of these CVs fall predominantly below the period gap (see \citealt{Knigge2006} and \citealt{Knigge2011}), but there were two within the gap (SSS1128-34 and possibly CSS2054-19) and six above it. The predominance of CVs with periods below the period gap in the CRTS dataset is not fully explained. \citet{Thorstensen2012} discussed this bias. They compared the cumulative distribution functions of the outburst amplitudes of the CRTS CVs and those in the survey region that were listed in RKcat  \citep{Ritter2003} that were not detected by the CRTS. They found that the CRTS shows a bias against low amplitude outbursts up to 6$\,$mag. CVs with larger outburst amplitudes spend more time above the 2$\,$mag cut-off limit that the CRTS employs and are thus more likely to be detected. Furthermore, they plotted the outburst amplitude versus the orbital period for CVs within the CRTS footprint (values from RKCat) and found a trend for short period DNe to have larger outbursts. Combined, these two findings indicate that there is a bias in the CRTS towards shorter period systems \citep{Thorstensen2012}.

Eight of the CVs in this sample were eclipsing systems. As mentioned in Section \ref{sec:observations}, the long-term CRTS light curves of the deeply eclipsing systems sometimes give indications of the eclipses. As each field is observed three times at 10-min intervals, it is possible that one or two of the observations are taken in eclipse. If it is sufficiently deep, these points can show up on the long-term light curve at a magnitude or more below the quiescent level. Of the five systems with eclipse depths of more than 0.9$\,$mag in our sample, four showed these characteristics.

A number of the CVs presented in this paper provide promising targets for more in-depth studies with larger telescopes - such as the eclipsing systems (CSS0116+09, CSS0438+00, CSS0449-18, CSS1221-10, SSS1224-41, SSS1340-35, SSS2003-28 and CSS2108-03), which through eclipse deconvolution can yield parameters such as the inclination and mass ratio. CSS1417-18 shows an unusually fast decline from outburst. The new orbital periods will also contribute towards population and evolutionary studies.

\section*{Acknowledgements}

We thank the anonymous referee for the suggestions and comments which helped to improve this paper.

The authors also gratefully acknowledge funding from the South African Square Kilometre Array Project, the Erasmus Mundus Programme SAPIENT, the National Research Foundation of South Africa (NRF), the Nederlandse Organisatie voor Wetenschappelijk Onderzoek (the Dutch Organisation for Science Research), the University of Cape Town, the National Astrophysics and Space Science Programme, the South African Astronomical Observatory (SAAO) and the Claude Leon Foundation Postdoctoral Fellowship program. Funding for the SHOC camera was provided by the NRF, specifically the Research Infrastructure Support Programme’s National Equipment Programme (grant UID \# 74428).

This research uses observations made at the SAAO and has also made use of NASA's Astrophysics Data System Bibliographic Services and of SDSS-III data. Funding for SDSS-III has been provided by the Alfred P. Sloan Foundation, the Participating Institutions (see http://www.sdss3.org), the National Science Foundation, and the U.S. Department of Energy Office of Science.

The CSS survey is funded by the National Aeronautics and Space Administration under Grant No. NNG05GF22G issued through the Science Mission Directorate Near-Earth Objects Observations Program. The CRTS survey is supported by the U.S. National Science Foundation under grants AST-0909182 and AST-1313422.

Thank you to Thuso Simon for his Markov Chain Monte Carlo code.


\label{lastpage}

\end{document}